\documentclass{moriond}

\def\Journal#1#2#3#4{{#1} {\bf #2}, #3 (#4)}

\def\PLB{{\em Phys. Lett.}  B}
\def\PRL{\em Phys. Rev. Lett.}
\def\PRD{{\em Phys. Rev.} D}
\def\PRC{{\em Phys. Rev.} C}

\def\be{\begin{equation}}
\def\ee{\end{equation}}
\def\bea{\begin{eqnarray}}
\def\eea{\end{eqnarray}}

\newcommand{\Photo}{}

\usepackage{xspace}
\newcommand{\pT}{\ensuremath{p_T}\xspace}

\begin{document}
\vspace*{4cm}
\title{RESULTS FROM PROTON-LEAD AND FIXED-TARGET COLLISIONS AT LHCB}

\author{ DANIELE MARANGOTTO on behalf of the LHCb Collaboration }

\address{INFN Milano \& Universit\`a degli Studi di Milano, Italy}

\maketitle\abstracts{The latest results obtained by the LHCb collaboration from proton-lead and proton-gas fixed-target collisions are presented. Results related to charm and beauty flavour hadron and quarkonia production in proton-lead collisions, being clean probes for Cold Nuclear Matter effects, are compatible to predictions based on initial-state effects. Indications of significant final-state effects in excited $\Upsilon(nS)$ resonances production are reported. The first cross-section measurement of $c\bar{c}$ and antiproton production in fixed-target proton-helium collisions at 100 GeV energy scale are also presented. More results will come from the latest Run 2 and future Run 3 data.}

\section{Introduction}
The LHCb detector, mainly designed to study flavour physics in proton-proton collisions, is the only LHC experiment fully instrumented in the forward pseudorapidity region $2<\eta<5$, providing complementary coverage with respect to the other LHC experiments. LHCb features excellent tracking performances, momentum resolution and particle identification capabilities.

Besides proton-proton collisions, LHCb also recorded proton-lead and lead-lead collisions. In asymmetric proton-lead collisions, LHCb is able to probe both forward and backward nucleon-nucleon centre-of-mass rapidity ($y^*$) by exchanging the proton and lead beam directions: LHCb covers the forward $1.5<y^*<4.0$ rapidity region when the proton beam goes towards the detector, and the backward $-5.0<y^*<2.5$ range when the lead beam is directed towards LHCb.

LHCb studied the production of charm and beauty quarks in proton-lead collisions, which is a clean probe for Cold Nuclear Matter (CNM) effects, comprising the modification of nuclear partonic distributions (nPDFs), initial-state radiation or coherent energy loss of the heavy quark, gluon saturation (described by the Colour Glass Condensate theory, CGC) and final-state hadronic rescatterings. The quantitative understanding of CNM effects is crucial for a correct interpretation of Quark Gluon Plasma (QGP) signatures in lead-lead collisions, where heavy quarks constitute special probes for QGP properties. LHCb provides measurements for CNM effects at low transverse momentum (\pT) and forward/backward rapidity, down to very small parton momentum fraction (Bj\"{o}rken $x \sim 10^{-5} - 10^{-6}$), where nPDFs are poorly constrained.

LHCb is also the only LHC experiment able to collect proton-gas fixed-target collisions thanks to its internal gas target, SMOG, which injects noble gases at $10^{-7} - 10^{-6}$ mbar pressures into the beampipe, and designed to perform precise luminosity measurements~\cite{Aaij:2014ida}. SMOG allows LHCb to explore fixed-target collisions at the unprecedented $\sqrt{s_{NN}}=100$ GeV nucleon-nucleon centre-of-mass energy scale.

\section{Observables and theoretical predictions}
Besides total and differential cross-sections, heavy quark production is studied by defining cross-section ratios as a function of \pT and $y^*$, less sensitive to systematic uncertainties and more easily comparable to theoretical predictions. The nuclear modification factor $R_{pPb}$ describe differences in heavy quark production between proton-lead and proton-proton collisions,
\begin{equation}
R_{pPb}(\pT,y^*) \equiv \frac{1}{208} \frac{d^2\sigma_{pPb}(\pT,y^*)/d\pT dy^*}{d^2\sigma_{pp}(\pT,y^*)/d\pT dy^*},
\end{equation}
which is sensitive to the effects of nuclear interactions.
The forward-to-backward ratio describes differences in heavy quark production for equal but opposite rapidity,
\begin{equation}
R_{FB}(\pT,y^*) \equiv \frac{d^2\sigma(\pT,+y^*)/d\pT dy^*}{d^2\sigma(\pT,-y^*)/d\pT dy^*}.
\end{equation}
LHCb also measured baryon-to-meson ratios, which are sensitive to the heavy quark hadronisation mechanism,
\begin{equation}
R_{B/M}(\pT,y^*) \equiv \frac{d^2\sigma_{B}(\pT,y^*)/d\pT dy^*}{d^2\sigma_{M}(\pT,y^*)/d\pT dy^*}.
\end{equation}

Measured ratios are compared to different theoretical predictions. Predictions based on initial-state (nPDFs) effects are obtained using the HELAC-Onia generator for collinear factorisation~\cite{Shao:2015vga} or a perturbative QCD calculation at fixed-order next-to-leading log (FONLL)~ \cite{Cacciari:1998it}, based on different nPDF sets: EPS09 at LO/NLO~\cite{Eskola:2009uj}, nCTEQ15~\cite{Kovarik:2015cma} and CT14NLO~\cite{Dulat:2015mca}. Where applicable, predictions from CGC effective field theory in dilute-dense approximation~\cite{Ducloue:2016pqr} and coherent energy loss from soft gluon radiation~\cite{Arleo:2012rs} are employed.

\section{Heavy quark production in proton-lead collisions}
LHCb studied the $J/\psi$ production at nucleon-nucleon centre-of-mass energy ofs $\sqrt{s_{NN}}=5.02$ and $8.16$ TeV energies, separating prompt quarkonia produced in the proton-lead interaction from displaced ones coming from $b$ quark decays~\cite{Aaij:2017cqq}. A strong suppression of the nuclear modification factor (down to 0.6) is observed for prompt $J/\psi$ at low \pT and forward $y^*$, while it is much less pronounced for backward rapidity. No significant suppression for $J/\psi$ from $b$ decays is seen. Predictions from nPDF effects, CGC and coherent energy loss follow the observed ratios, but on average they tend to overestimate data, indicating less anti-shadowing than expected. The forward-to-backward ratio is found to rise with \pT, especially for prompt $J/\psi$, while a flat rapidity dependence is seen.

The production of the $\Lambda^+_c$ charm baryon is studied at $\sqrt{s_{NN}}=5.02$ TeV~\cite{Aaij:2018iyy}. The forward-to-backward ratio is found consistent with HELAC-Onia predictions. The baryon-to-meson ratio $R_{\Lambda^+_c/D^0}$ agrees with results from proton-proton data but for forward rapidity and high \pT, where the $\Lambda^+_c$ baryon production seems suppressed.

LHCb studied open beauty hadron production at $\sqrt{s_{NN}}=8.16$ TeV~\cite{Aaij:2019lkm}, reconstructing beauty hadrons in the exclusive decay modes $B^+\to \bar{D}^0\pi^+$, $B^+\to J/\psi K^+$, $B^0\to D^-\pi^+$ and $\Lambda^0_b\to \Lambda^+_c\pi^-$. This is the first measurement in nuclear collisions down to very low \pT. The measured nuclear modification factor confirms the suppression pattern seen in $J/\psi$ from $b$ decays production and the $\Lambda^0_b$-to-meson ratio is consistent with proton-proton data results.

\begin{figure}
\centering
\includegraphics[scale=0.24]{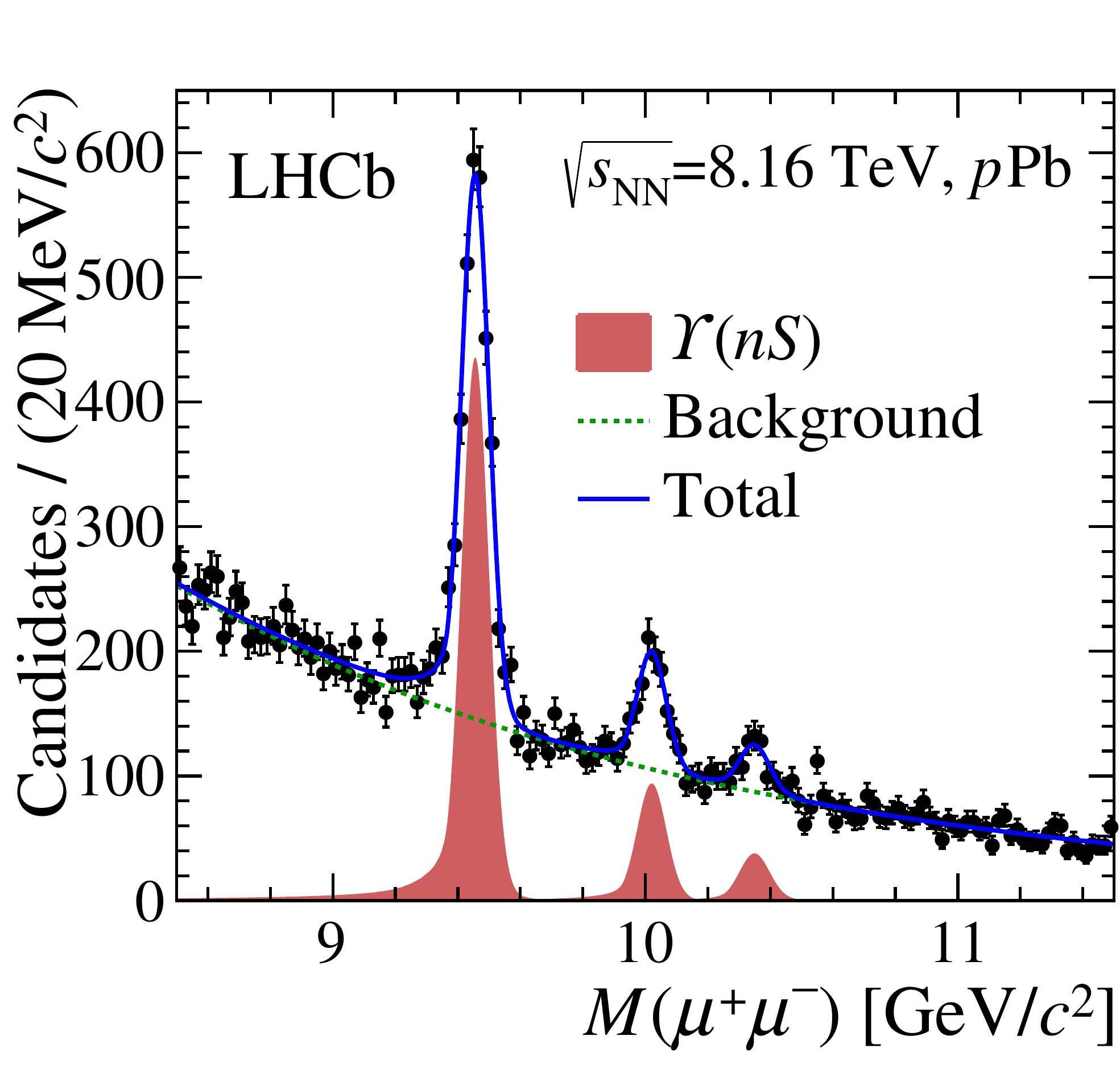}
\includegraphics[scale=0.24]{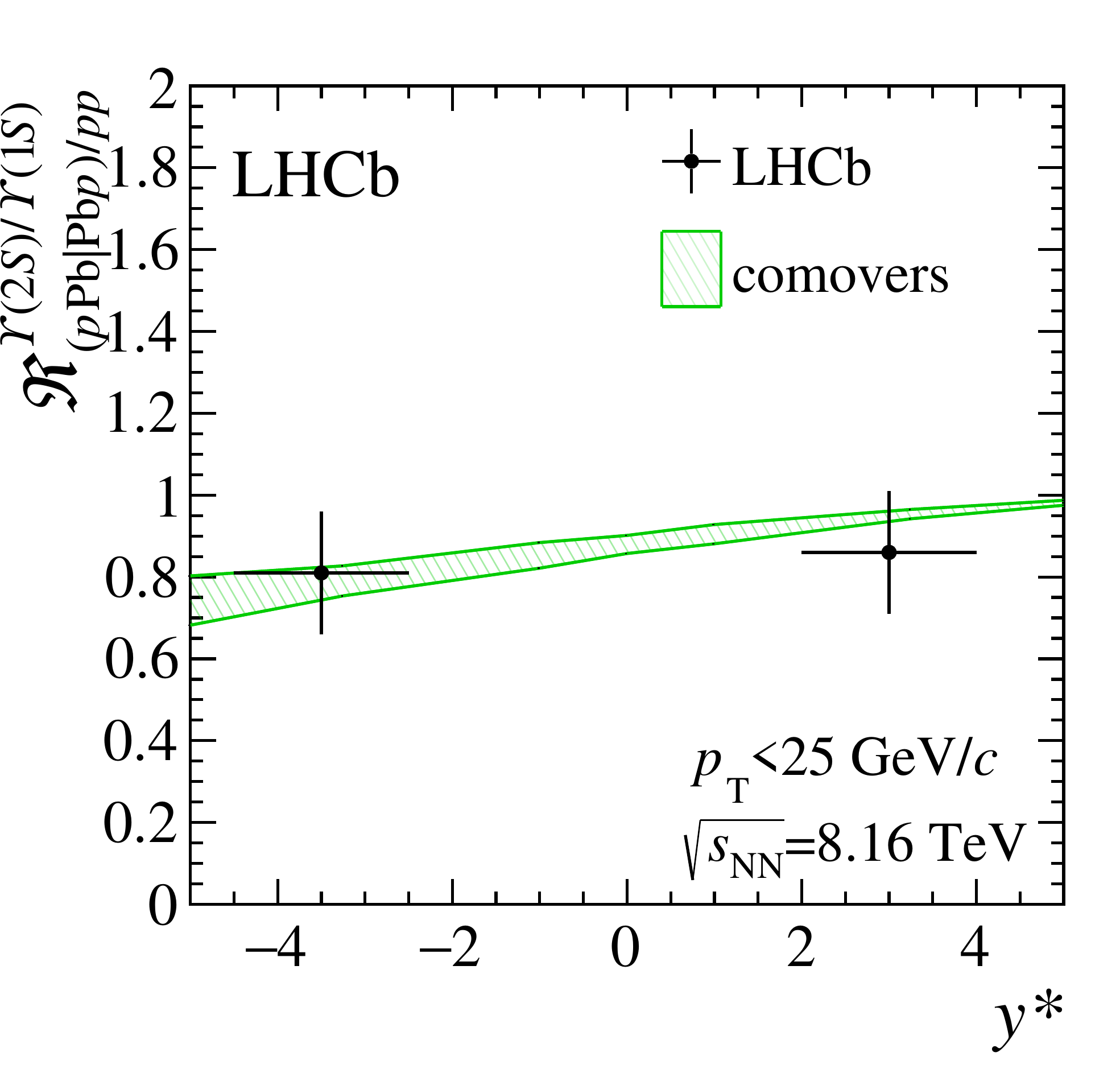}
\includegraphics[scale=0.24]{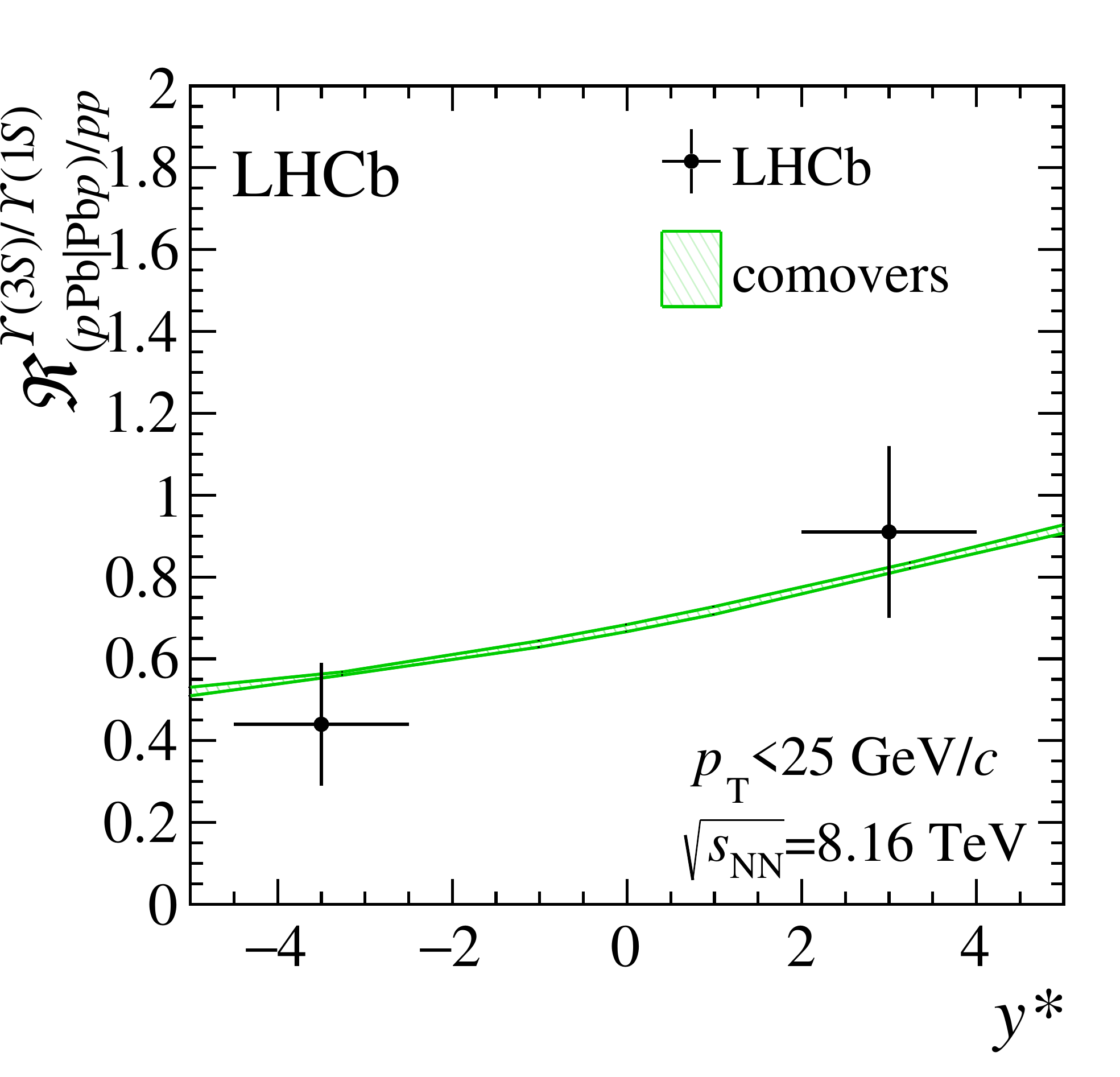}
\caption{(left) Reconstructed $\Upsilon(nS)\to\mu^+\mu^-$ invariant mass and nuclear modification factor for (center) $\Upsilon(2S)$ and (right) $\Upsilon(3S)$ states normalized to the $\Upsilon(1S)$ one.}
\label{fig:pPb_upsilon}
\end{figure}

The production of the three $\Upsilon(nS)$ states is studied at $\sqrt{s_{NN}}=8.16$ TeV~\cite{Aaij:2018scz}. Thanks to the performances of the LHCb detector the three $\Upsilon(nS)$ are clearly resolved in the reconstructed $\mu^+\mu^-$ invariant mass. The suppression of nuclear modification factors for excited $\Upsilon(2S)$, $\Upsilon(3S)$ resonances is expected to be larger than the $\Upsilon(1S)$ state according to the ``comover model'' final-state effects~\cite{Ferreiro:2018wbd}, especially for backward rapidities. This model predicts the production of excited $\Upsilon(nS)$ resonances to be reduced due to the interaction of comoving particles, a possible explanation for the $\Upsilon(nS)$ suppression observed in lead-lead collisions. The measured nuclear modification factors show indeed a larger suppression for excited $\Upsilon(nS)$ quarkonia states, compatible with the ``comover'' model predictions, Fig.~\ref{fig:pPb_upsilon}. 

\section{Fixed-target p-gas collisions studies}
\begin{figure}
\centering
\includegraphics[scale=0.27]{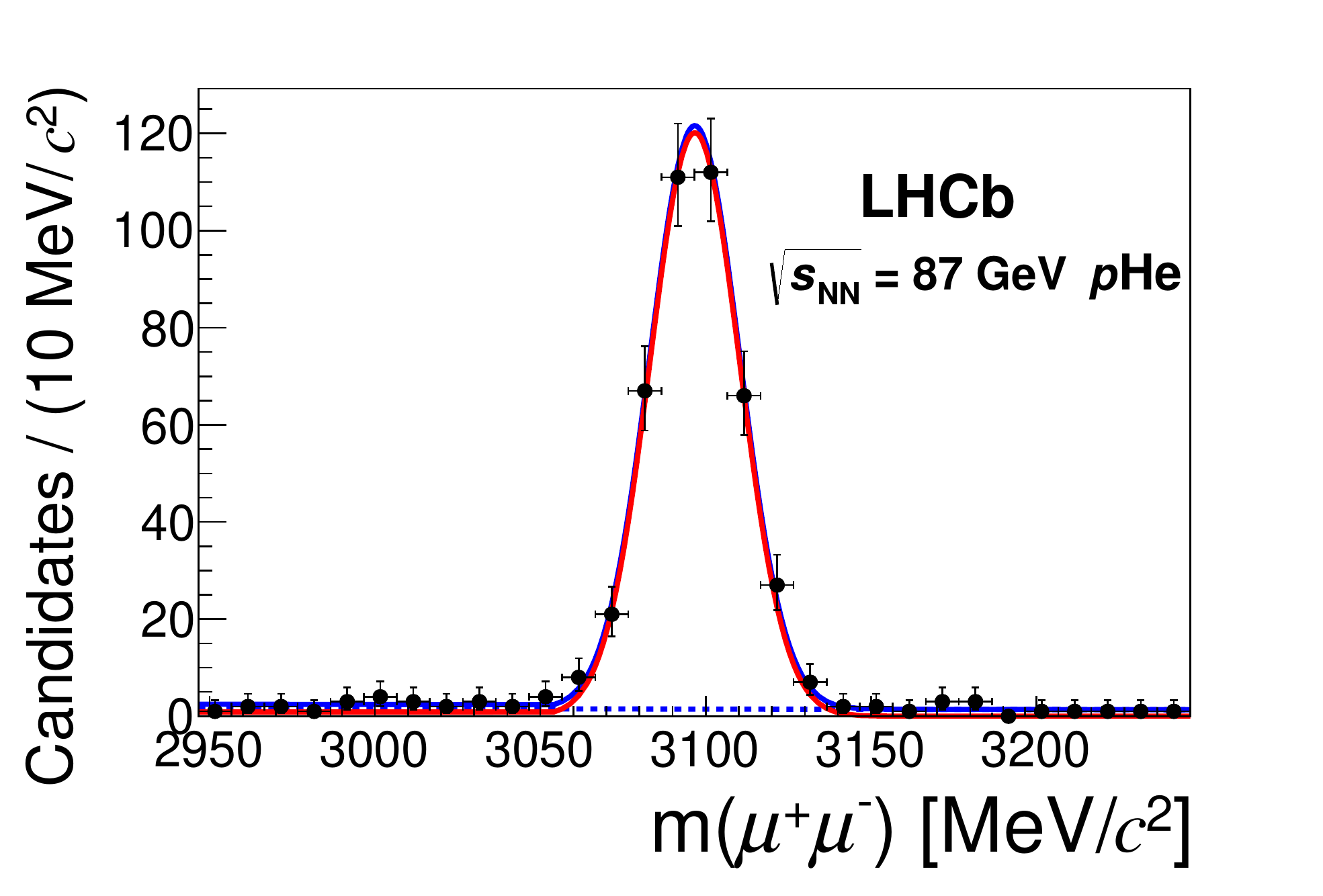}
\includegraphics[scale=0.25]{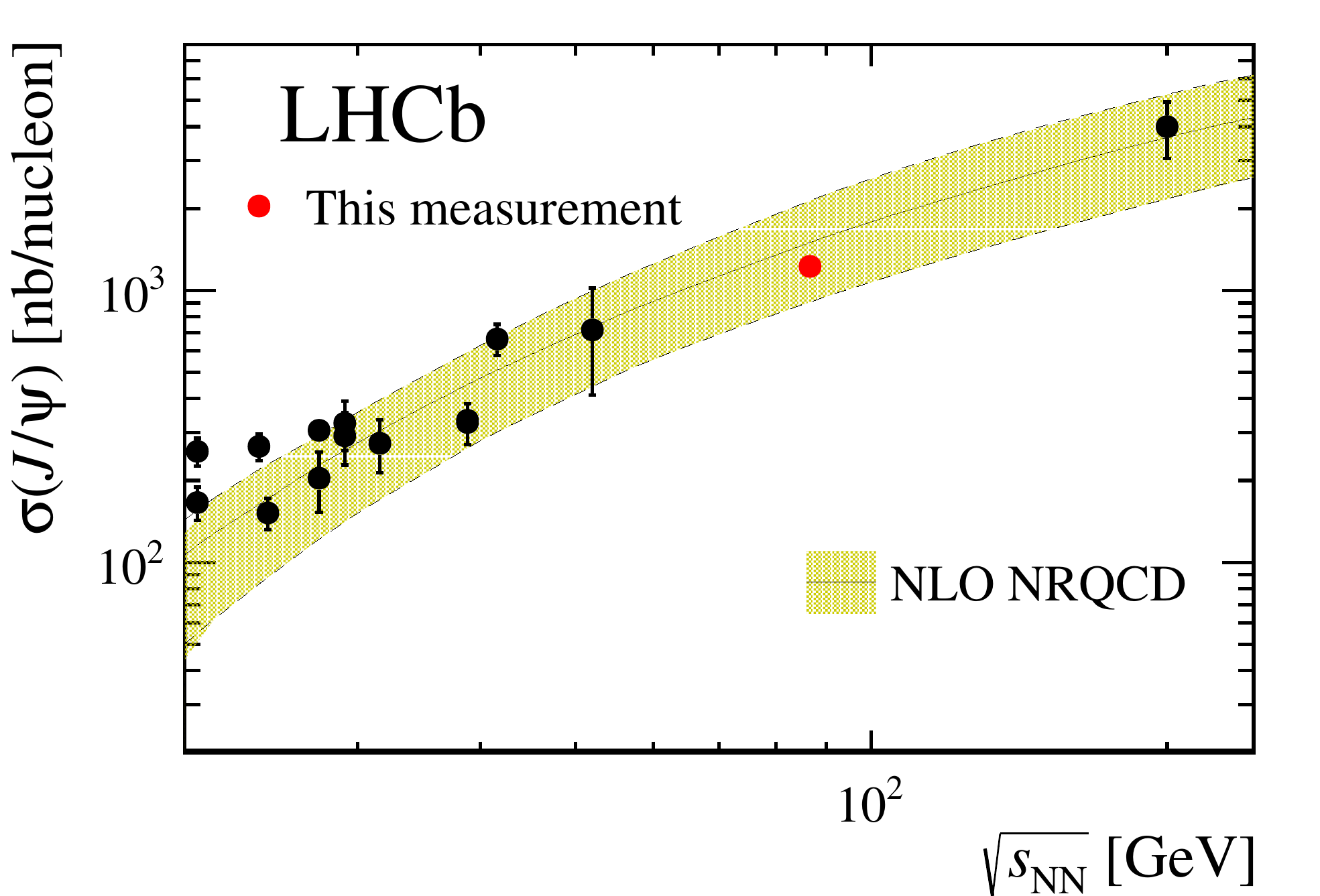}
\includegraphics[scale=0.25]{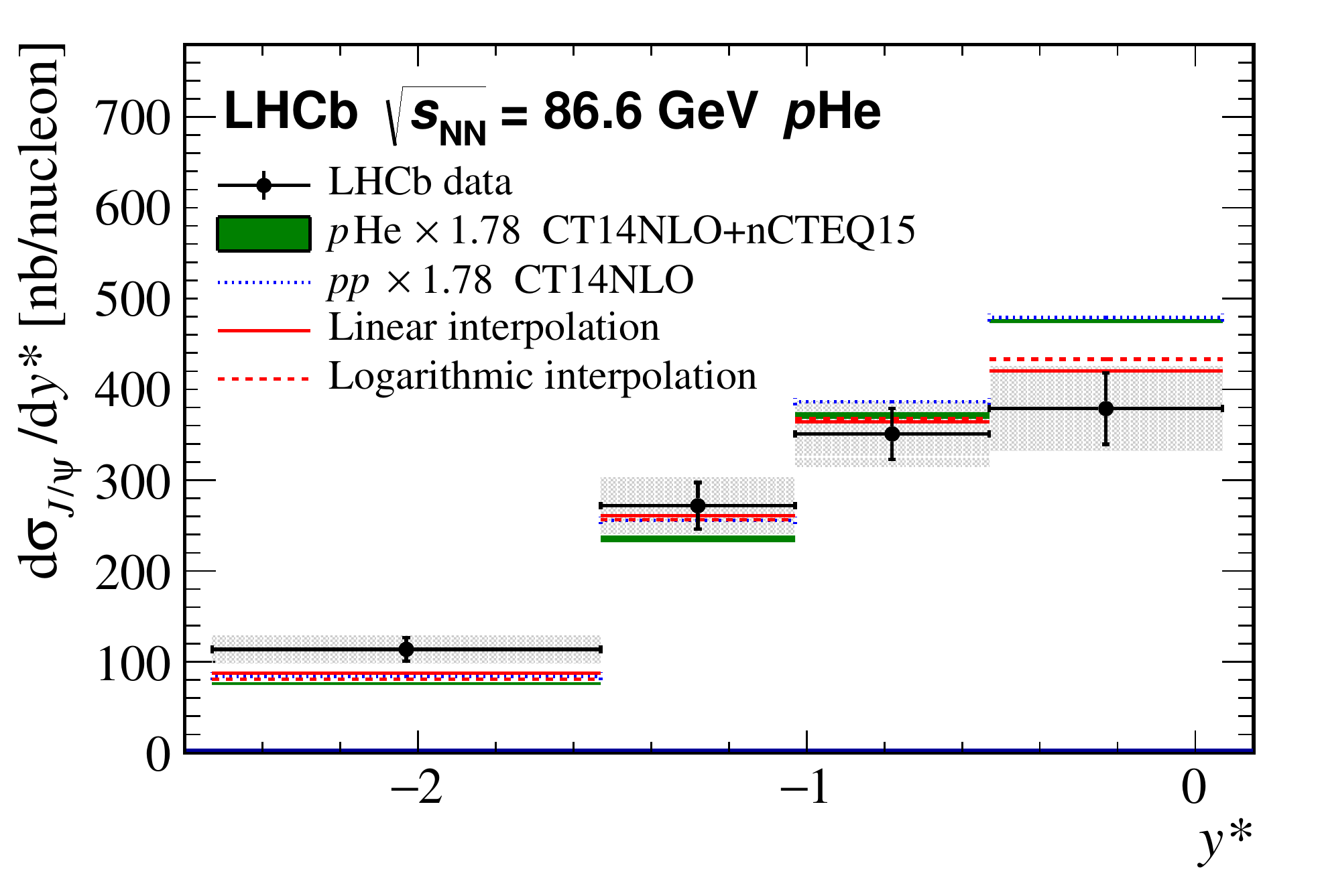}
\caption{(left) Reconstructed $J/\psi\to\mu^+\mu^-$ invariant mass and (center) total and (right) differential $J/\psi$ cross-sections, obtained from the proton-helium dataset at $\sqrt{s_{NN}}=87$ GeV.}
\label{fig:FT_charm}
\end{figure}

LHCb collected the first charm ($J/\psi$ and $D^0$) samples in proton-argon and proton-helium collisions at $\sqrt{s_{NN}}=110$ and $87$ GeV energy, respectively~\cite{Aaij:2018ogq}. The first $c\bar{c}$ cross-section measurement is performed using the proton-helium dataset, Fig.~\ref{fig:FT_charm}, for which the integrated luminosity was determined from the yield of elastically scattered electrons off the target helium atoms~\cite{Aaij:2018svt}. Predictions from HELAC-Onia generator follow the shape of the measured differential cross-sections but underestimate the total by a factor 1.5-1.8. Results are also compared to linear and logarithmic interpolations from cross-section results at the closest energy available. From the measured differential cross-section it is possible to infer that there is no evidence for a significant $c$ quark content in the nPDF functions in the kinematic region probed by LHCb.~\cite{Pumplin:2007wg}.

Space-based high-energy cosmic ray experiments AMS-02 and PAMELA have observed an excess of the antiproton-to-proton ($\bar{p}/p$) ratio at the 100 GeV energy scale~\cite{Aguilar:2016kjl}. An enhancement of the $\bar{p}/p$ ratio can be an indirect signal of dark matter, however the significance of the result is limited by the large uncertainties related to antiproton production predictions. In particular, the antiproton production cross-sections in proton-hydrogen and proton-helium collisions are poorly known. LHCb performed the first measurement of antiproton production in proton-helium collisions at $\sqrt{s_{NN}}=110$ GeV~\cite{Aaij:2018svt}, providing precious information for more precise $\bar{p}/p$ ratio predictions. The measured total and differential cross-sections have been compared to predictions from different generators~\cite{Pierog:2013ria}. The absolute antiproton yield is found to be underestimated by predictions up to a factor two, a result tending to disfavour the hypothesis of a dark matter origin of the measured $\bar{p}/p$ ratio excess in cosmic rays. Generators are able to reproduce the shape of the measured proton-helium differential cross-section, but for low momentum antiproton production is found to be lower than expected.

\section{Prospects}
The presented studies are just the beginning of the LHCb physics program with heavy ion and fixed-target collisions. Much more analyses are to come, like Drell-Yan and vector boson production studies, the investigation of more quarkonia states and dihadron correlations, Bose-Einstein condensates and flow studies. The latest LHC Run 2 samples of proton-neon, lead-lead and lead-neon collisions, the LHCb datasets with higher statistics in their respective configurations, are being analysed. At least a factor ten more integrated luminosity of proton-lead and lead-lead collisions is planned for the LHC Run 3 data-taking, to be recorded with an upgraded LHCb detector. The LHCb Collaboration is considering the installation of an upgraded gas injection system, SMOG2, aiming at increasing the injected gas pressure, therefore the luminosity of fixed-target collisions, by a factor 100. SMOG2 would also allow to inject more gas species (not only noble gases) with a precise density control. The LHCb Collaboration is also considering a Phase II upgrade of the LHCb detector for the LHC Run 5, for which a detailed physics case has been presented~\cite{Bediaga:2018lhg}.

A rich physics program with heavy ions and fixed-target is ahead for the LHCb experiment.

%\begin{figure}
%\begin{minipage}{0.33\linewidth}
%\centerline{\includegraphics[width=0.7\linewidth,draft=true]{figexamp}}
%\end{minipage}
%\hfill
%\begin{minipage}{0.32\linewidth}
%\centerline{\includegraphics[width=0.7\linewidth]{figexamp}}
%\end{minipage}
%\hfill
%\begin{minipage}{0.32\linewidth}
%\centerline{\includegraphics[angle=-45,width=0.7\linewidth]{figexamp}}
%\end{minipage}
%\caption[]{same figure with draft option (left), normal (center) and rotated (right)}
%\label{fig:radish}
%\end{figure}

%\section*{Acknowledgments}
%
%This is where one places acknowledgments for funding bodies etc.
%Note that there are no section numbers for the Acknowledgments, Appendix
%or References.

\section*{References}

%\bibliographystyle{plain}
%\bibliography{biblio}

\end{document}